\documentclass{pspum-l}
\usepackage{psfig}

\begin{document} 
 
\title[Self-Organized Percolation Power Laws without Fractal
Geometry]{Self-Organized Percolation Power Laws with and without
Fractal Geometry in the Etching of Random Solids}
 
\author[A.~Desolneux]{Agn\`es Desolneux$^\dag$ $^1$}
\author[B.~Sapoval]{Bernard Sapoval$^\P$ $^{1,2}$}
\author[A.~Baldassarri]{Andrea Baldassarri$^\S$ $^{1-3}$} 

\address{$^1$ Centre de Math\'ematiques et de leurs Applications,
CNRS, Ecole Normale Sup\'erieure, 94235 Cachan, France}

\address{$^\dag$  Agnes.Desolneux@cmla.ens-cachan.fr}

\address{$^2$ Laboratoire de Physique de la Mati\`ere Condens\'ee,
CNRS, Ecole Polytechnique,  91128 Palaiseau, France}

\address{$^\P$  Bernard.Sapoval@polytechnique.fr}

\address{$^3$ INFM, Unit\'a di Roma, Dipartimento di Fisica,
Universit\'a ``La Sapienza'', P.le A.Moro 2, 00185 Roma, Italy}

\address{$^\S$  Andrea.Baldassarri@roma1.infn.it}


\date{\today }
 
\begin{abstract} 
 
Classically, percolation critical exponents are linked to the power
laws that characterize percolation cluster fractal properties. It is
found here that the gradient percolation power laws are conserved even
for extreme gradient values for which the frontier of the infinite
cluster is no more fractal. 

In particular the exponent $7/4$ which was recently demonstrated to be
the exact value for the dimension of the so-called "hull" or external
perimeter of the incipient percolation cluster, controls the width and
length of gradient percolation frontiers whatever the gradient
magnitude.

This behavior is extended to previous model studies of etching by a
finite volume of etching solution in contact with a disordered
solid. In such a model, the dynamics stop spontaneously on an
equilibrium self-similar surface similar to the fractal frontier of
gradient percolation.  It is shown that the
power laws describing the system geometry involves also the fractal
dimension of the percolation hull, whatever the value of the
dynamically generated gradient, i.e. even for a non-fractal frontier.

The comparison between numerical results and the exact
results that can be obtained analytically for extreme values of the
gradient suggests that there exist a unique power law valid from the
smallest possible scale up to infinity. These results suggest the
possible existence of an underlying conservation law, relating
the length and the statistical width of percolation gradient frontiers.

\end{abstract} 

\maketitle

\specialsection{Introduction: Gradient percolation built by diffusion
or by etching} 

Spreading of objects in space with a gradient of
occupation probability is most common. From chemical composition
gradients to the distribution of plants which depend on their solar
exposure, probability gradients exist in many inhomogeneous
systems. In fact inhomogeneity is a rule in nature whereas most of the
systems that physicists are studying are homogeneous as they are
thought to be more simple to understand. In particular phase
transitions or critical phenomena are studied in that framework, the
simplest being percolation transition \cite{Stauffer}. In this work,
we study a different situation, that of an inhomogeneous system, the
gradient percolation situation.

First we study a paradigmatic situation of such an inhomogeneous system, the
Gradient Percolation {\bf GP} model \cite{Sapo1,gp}.
We show how the scaling behavior, which  usually
characterizes criticality, extends down to minimal possible
scales, i.e. the extreme gradient regime. 
In this regime, exact analytical analysis of the model is possible and given here. 

Moreover this analysis can be extended to other models, as the etching
gradient percolation dynamical model {\bf EGP}, which has recently been shown to
belong to the same universality class as GP.  In this case,
the extreme gradient analysis allows to correctly recognize such a
universality (and its limitations) {\it without}  the explicit knowledge of
the underlying dynamically generated gradient.

\begin{figure}[tbp] 
\centerline{\psfig{figure=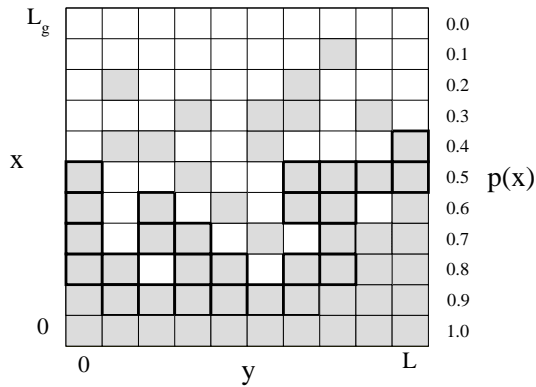,width=8cm,angle=-90}}
\caption{The Gradient Percolation Model. Each row has an occupation
probability $p(x)$ ranging from $1$ at the bottom to $0$ on the
top.  A site with coordinates $(x,y)$ is occupied with probability $p(x)$. The occupied and
empty sites are represented respectively in grey and white. Apart from
isolated islands and lakes, grey and white sites form two distinct
connected regions.  The marked sites (with thick borders) are the
connected front of the occupied percolating cluster.\label{gp-sketch}}
\end{figure} 

\begin{figure}[tbp] 
\begin{center} 
\centerline{\psfig{figure=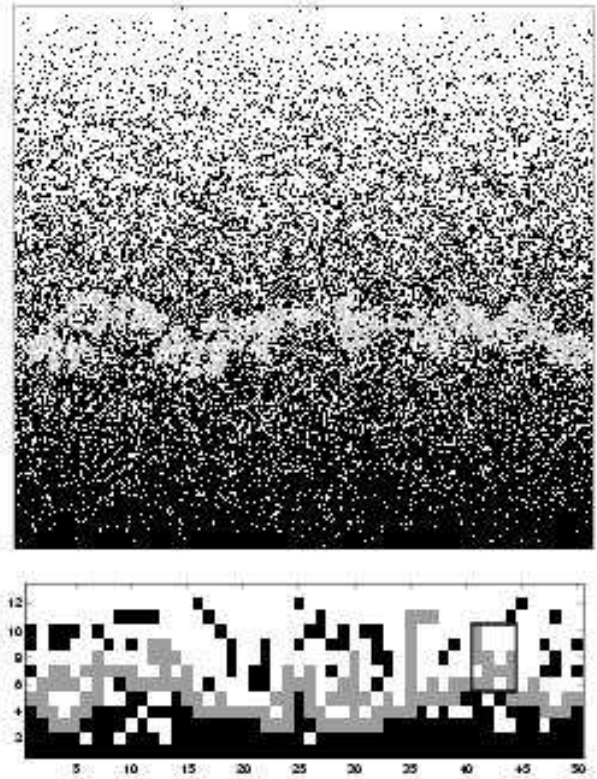,width=8cm}}
\end{center} 
\caption{Gradient percolation (GP) front. Particles are distributed at
random with probability $p(x)=1-x/L_g$. The occupied sites are in
black and the last line of connected occupied sites is the gradient
percolation front shown in light gray. This situation corresponds also
to a diffusion situation where the front is called the diffusion
front. Top: $L=L_g=500$. Bottom: $L=50$ and $L_g=12$. The black window
has an horizontal width equal to $\sigma_f$. It contains approximately
$L_g$ points.}
\label{fig:G.P.} 
\end{figure}

\subsection*{Gradient Percolation GP}
\subsubsection*{GP Definition of Gradient Percolation}
The gradient percolation (GP) situation was first introduced in the
study of diffusion fronts which exists for instance as the consequence
of a soldering process \cite{Sapo1,gp}. It is defined in
Fig.~\ref{gp-sketch} and examples are shown in
Fig.~\ref{fig:G.P.}. The figures give examples of a random
distribution of points on a lattice with a linear gradient of
concentration in the vertical direction.

It is a 2D square lattice of
size $L_g\times L$, where each point $(x,y)$ is occupied with
probability $p(x)=1-x/L_g$ ($x$ being the vertical direction in the
figure). In gradient percolation there is always an infinite cluster
of occupied sites as there is a region where $p$ is larger than the
standard percolation {\bf SP} threshold $p_c$. There is also an infinite
cluster of empty sites as there is a region where $p$ is smaller than
$p_c$. The object of interest is the GP front, the external limit
(or frontier) of the infinite occupied cluster. 

Its precise definition depends on the lattice geometry:
\begin{itemize}
\item Triangular Lattice.
The GP front is the connected set of sites belonging to the occupied
infinite cluster which are first nearest neighbours to sites of the
infinite empty cluster.
\item Square Lattice. 
The GP front is the connected set of sites belonging to the occupied
infinite cluster which are first or second nearest neighbours with
sites of the infinite empty cluster, itself defined through a first
and second neighbours connection.
\end{itemize}
The square lattice case is shown in grey in Fig.~\ref{fig:G.P.}. This
front is a random object with an average position $x_f$, a statistical
width $\sigma_f$ and a total length $N_f$. In so far that the GP front
and the SP external perimeter (often called hull) have the same
geometry their fractal dimension was first conjectured to be exactly
equal to 7/4 in \cite{Sapo1}. This result was then demonstrated
heuristically by Saleur and Duplantier \cite{Saleur87} and very
recently it was proved mathematically by Smirnov and Werner
\cite{Smirnov}.
 
\subsection*{Etching Gradient Percolation EGP}

The same type of fractal geometry has been recently found in a very
different physical situation. It first appeared in experiments
\cite{Balazs}. It was interpreted in a model in which an etching solution is
in contact with the initially flat surface of a disordered solid and
starts to corrode its weakest regions. The solid surface then gets
``harder'' but at the same time new regions are discovered which
contain weak elements. Often the corrosive power of the solution is
proportional to an etchant concentration and if the etchant is
consumed in the reaction, the corrosive power of a {\it finite
volume} of solution decreases during the time evolution of the
process. As the solid surface gets harder and harder, and the
corroding power  gets weaker and weaker, the corrosion
process stops spontaneously in a finite time interval.

\begin{figure}[tbp] 
\centerline{ 
\psfig{figure=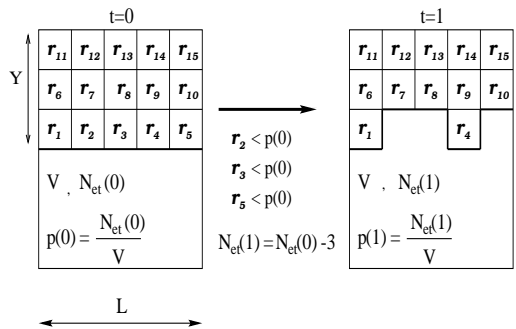,width=7cm,angle=-90}
}
\caption{ Sketch of the etching dynamics in a square lattice: sites
$2,3,5$ are etched at the first time-step as their resistances are
smaller than $p(0)$. Consequently the number of etchant particles in
the solution decreases by $3$ units. At $t=1$, the new interface sites
than can possibly be etched are $7,8,10$ if the solution can etch only
the first nearest neighbours.  If the solution can also etch second
nearest neighbours in a diagonal direction the whole second layer can
possibly be etched. }
\label{etching-sketch} 
\end{figure} 

\subsubsection*{The EGP model definition} 
We first recall the two-dimensional etching model introduced in
\cite{sapo-etch}.  Its schematic is shown in Fig.~\ref{etching-sketch}:
\begin{itemize} 
\item  A 2D random solid is represented as a site lattice (triangular or square), of 
linear width $L$ and, eventually, infinite depth.
\item  A random number $r_{i}\in [0,1]$ (extracted from the flat probability 
density function $\pi _{0}(r)=1$ for $r\in [0,1]$) is assigned to each
solid site $i$, representing its resistance to the etching by the
solution. $r_{i}$ does not depend on time (quenched disorder), or on
the site environment.
\item  The etching solution has a volume $V$ and is initially in 
contact with the solid through the bottom boundary (see
Fig.~\ref{etching-sketch}).  It contains an initial number $N_{et}(0)$ of
dissolved etchant molecules.
\end{itemize}

Consequently, the initial concentration $C(0)$ of etchant in the
solution is given by: $C(0)=N_{et}(0)/V$. Calling $N_{et}(t)$ the
number of etchant molecules at time $t$, $C(t)=N_{et}(t)/V$. At each
time-step, the ``etching power'' of the solution (i.e. the average
``force'' exerted by the solution on a solid surface particle) is
supposed to be proportional to $C(t)$ : $p(t)=\Gamma C(t)$.  Hereafter
the assumption $\Gamma =1$ is made, without loss of generality. It
implies $C(t)\equiv p(t)$. At time-step $t$, all the interface sites
with $r_{i}<p(t)$ are dissolved and a particle of etchant is consumed
for each of these corroded solid sites.
 
\begin{figure}[tbp] 
\centerline{\psfig{figure=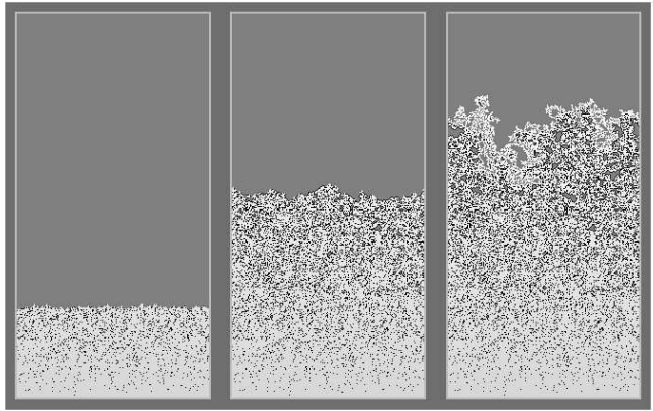,width=10cm,angle=-90}}
\caption{Etching process for two intermediate times
and final equilibrium situation. The solid is shown in grey, the
solution in white, and the finite size solid clusters that are
detached by the etching in black.  The final solid surface is found to
be fractal with dimension $7/4$ up to a characteristic scale $\sigma_f$.}
\label{etching-pictures} 
\end{figure}

The etching evolution is depicted in Fig.~\ref{etching-pictures}. A
brief review of the known results, will be given in the next section.

The structure of this paper is the following: in the next section we
recall definitions and known results for fractal "weak gradient"
cases. Then we consider the extreme gradient case for GP. In that case
a few exact results are given. They are discussed in the spirit of
deciding whether or not there exists power laws for GP that are valid
from the smallest scale to infinity.

They are then compared with numerical results
obtained for small size systems.
Next the EGP is studied numerically for large gradients. The
results are then extrapolated to large (fractal) systems and a
remarkable agreement is found between extrapolated values and
numerical values. Gradient percolation results are then compared with
etching gradient percolation. In the conclusion we discuss to what
extent the numerical results give evidence that there exists a new conservation
law in diffusion or gradient percolation, this law being valid 
from the smallest scale to infinity.

\specialsection{GP and EGP Known Results}
\subsection{Gradient Percolation (GP): Known Results}
The early GP studies were focussed at finding its relations with
standard percolation. Let us first recall the definitions. For $0\leq
x\leq L_g$, $n_f(x)$ is the mean number of points of the front lying
on the line $x$ per unit horizontal length. It measures the front
density at distance $x$. The length $N_f$, the position $x_f$ and the
width $\sigma_f$ of the front are then defined in terms of 
$n_f(x)$ by $$N_f=L\sum_{x=0}^{L_g} n_f(x) , \hspace{0.2cm} x_f =
\frac{\sum_{x=0}^{L_g} x n_f(x)}{\sum_{x=0}^{L_g} n_f(x)} , $$ $$
{\mathrm{and}} \hspace{0.2cm} \sigma_f^2=
\frac{\sum_{x=0}^{L_g}(x-x_f)^2 n_f(x)}{\sum_{x=0}^{L_g} n_f(x)} .$$

It was found that the mean front was located at a distance where the
density of occupation was very close to $p_c$ or $p(x_f) \simeq
p_c$. This was verified numerically with such precision that the
gradient percolation method is now often used to compute percolation
thresholds \cite{Rosso1,Rosso2,Ziff87,Quintinilla1,Quintinilla2}. It
was also found that:

\begin{enumerate} 
\item The width $\sigma_f$ depends on $L_g$ through a power law $\sigma_f 
\propto (L_g)^{\nu/(1+\nu)}$ where $\nu=4/3$ is the correlation length 
exponent \cite{Stauffer} in dimension $d=2$ so that $\sigma_f =
(L_g)^{4/7}$.  The width $\sigma_f$ was also shown to be a percolation
correlation length.
\item Secondly it was found that the front was fractal with a dimension $D_f$, 
numerically determined, close to 1.75. The front length followed a
power law $N_f \propto (L_g)^{\alpha_N}$ with $\alpha_N =
(D_f-1)\nu/(1+\nu) $.
\item But also, it was numerically observed that the sum of these two 
exponents was very close to 1. If true, this meant that $ \nu/(1+\nu)
+ (D_f-1)\nu/(1+\nu) =1$ or $D_f = 1+1/\nu$ = 7/4.  This is how it was
conjectured in \cite{Sapo1} that $D_f = 7/4$.

\end{enumerate} 

\noindent 
 
In that sense the ordinary GP power laws  were
thought to be linked to the SP exponent $\nu$ and to the fractality
of the percolation cluster hull. Up to now, these facts were
considered to be strictly valid only in the large system limit.
 
However, if true, and we know now that 7/4 is the exact value, there
follows an intriguing relation, namely $\sigma_f^{D_f}$ is exactly
proportional to $L_g$. This means that the number of surface
particles within the correlation length is exactly proportional to
$L_g$. This is particularly striking for diffusion fronts. Diffusion
of particles from a source results in a concentration gradient and an
associated GP situation. In that frame, the above result means that,
if $L_g$ particles have diffused on a vertical row, there is on
average the same (or a constant fraction of) number of particles on
the correlated surface (surface content of a box with a lateral size
equal to the statistical width). This fact seems a priori to have
nothing to do with scaling, percolation and the thermodynamic
limit. From this point of view it is possibly the consequence of a
conservation law and if such a conservation exists, it should apply
also for extreme gradients corresponding to $L_g$ of a few units. 

In particular it should apply to the very extreme $L_g=1$, $2$ and $3$
for which exact values of $x_f$, $N_f$ and $\sigma_f$ can be
calculated analytically. The corresponding results are the subject of
section 3.

\subsection{Etching Gradient Percolation (EGP): Known Results}

\begin{figure}[tbp] 
\centerline{\psfig{figure=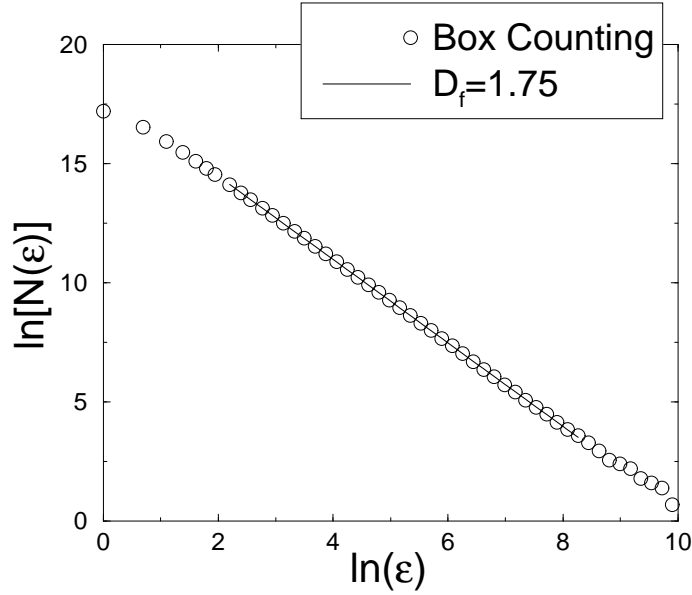,width=9cm}}
\caption{Box-counting determination of the fractal dimension $D_f$ of
the corrosion front ($\epsilon$ is the linear size of the box,
$N(\epsilon)$ the number of boxes containing at least one point of the
front). The value of $D_f=1.753\pm 0.005$ is found fit for values
of $\epsilon$ ranging from a few lattice distances to the front width
$\sigma_f$ (in this case $\sigma_f \approx 3000$, i.e. $\ln(\sigma_f) \approx
8 $).
\label{dimension}} 
\end{figure} 
 
Let us call $n(t)$ the number of dissolved solid sites at time-step
$t$. One can express several quantities through
$n(t)$, or its time-integral $N(t)$, that is the total number of
corroded solid sites up to time $t$.
The number of etchant particles in the liquid will decrease as:
\begin{equation} 
N_{et}(t+1)=N_{et}(t)-n(t)=N_{et}(0)-N(t)\,, \label{eq1}
\end{equation} 
and consequently the etching power of the solution is:
\begin{equation} 
p(t+1)=p(t)-\frac{n(t)}{V}=p(0)-\frac{N(t)}V\,.  \label{eq2}
\end{equation} 
Note that, as $p(t+1)<p(t)$, a site having resisted to etching at a
certain time-step will resist forever.  Consequently, the part of the
solid surface which can be etched at time-step $t+1$ is restricted to
the sites which have been just uncovered by the etching process at
time $t$. We call this subset of surface the ``active'' part of the
surface. After a given time-step, all the solid sites which have been
previously explored by the solution are ``passive'' forever. However it may happen that ``passive'' sites are disconnected from the
bulk at a later time-step if they are connected to the solid by weak
sites.

The model reproduces qualitatively the same phenomenology observed
experimentally \cite{Balazs}. The dynamical evolution can be divided
into two different regimes:
\begin{enumerate} 
\item In the first (smooth) regime, the corrosion is well directed while 
the front becomes progressively rougher and rougher. In our model
this regime does not depend on the details of the discretization, 
or on the fundamental geometrical features of the
lattice, like the embedding space dimension or the lattice
coordination number.
\item In the second regime, the correlations revealed by the 
hardening process become important: the dynamics becomes locally
isotropic, generating a fractal front.  This corresponds to a critical
regime, directly related to the percolation transition on the
same lattice.
\end{enumerate}

In the case of etching the same results as for GP were found on
the final front geometry. The fractal dimension $D_{f}$ of the etching
front was measured (up to the scale $\sigma_f $) using the box-counting
\cite{box-counting} algorithm.  In this way $D_{f}=1.753\pm 0.005$ is
measured (see Fig.~\ref{dimension}) very close to the value $7/4$ of GP. 

\begin{figure}[tbp] 
\begin{center}
\centerline{ 
\psfig{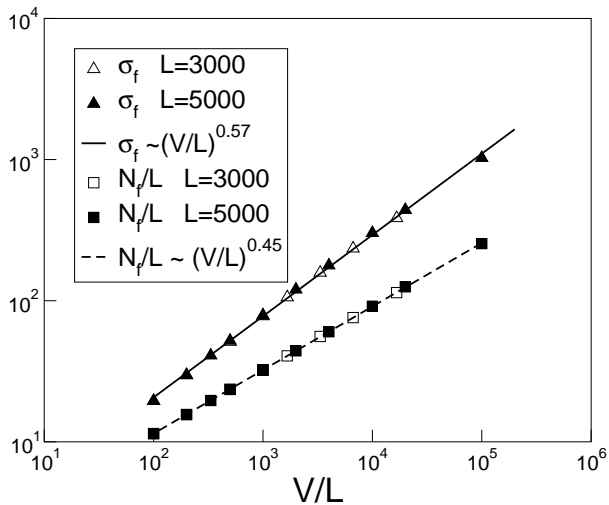}
}
\end{center} 
\caption{Behavior of $\sigma_f$ and of the average number of sites per
column $N_f/L$ as a function of $V/L$ for several sample sizes
(several values of $L$).
\label{scaling-fig}} 
\end{figure} 

In fact it was shown~\cite{GBS} that in EGP the length $V/L$, ratio of
the solution volume by the sample width, plays the role of $L_g$.  The
width and length of the front were found to follow the same power laws
as in GP, as shown in Fig.~\ref{scaling-fig}.  In particular
$\sigma_f$ scales as $(V/L)^{\alpha_\sigma}$, with
$\alpha_\sigma\approx 0.57$, while $N_f/L$ scales as
$(V/L)^{\alpha_N}$, with an observed value for $\alpha_N\approx 0.45$.

\begin{figure}[tbp] 
\centerline{ 
\psfig{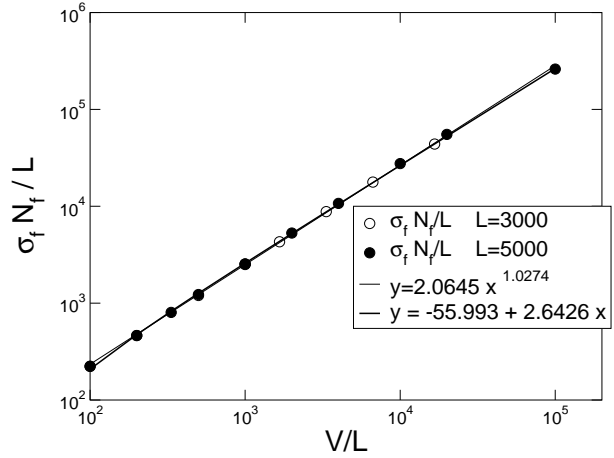}
}
\caption{Behavior of $\sigma_f N_f/L$ as a function of $V/L$. Data are
fitted with a power law, resulting in an exponent very close to one,
and with a linear law, resulting in comparable results.
\label{scaling-fig2}} 
\end{figure}

The scaling relation $\alpha_\sigma+\alpha_N=1$ seems also fairly
obeyed, but a direct inspection give an even better result, as shown in
Fig.~\ref{scaling-fig2}, where $\sigma_f N_f/L$ is plotted as a
function of $V/L$. A power law fit gives an exponent very close to one,
and correspondingly, a linear fit gives equivalently good results.
 
As the same relations are verified by EGP, there should also exist a
conservation law for that case. This law would stipulate that the
amount of correlated final front is proportional to the ratio of the
solution volume divided by the lateral size of the sample, in other
words to the depth of the solution. Formulated in this manner there is
no reason why this property should not be verified for very small
solution volumes, as shown in the next section.

\specialsection{Extreme Gradients: fractal exponents without fractal geometry}

In this section we will show how, even in the case of extreme gradient
(hence in absence of any fractal structure) one recovers the above
power laws.

\subsection{Exact calculations for extreme GP} 

We present a detailed description of some exact results
obtained for $L_g=2,3$, hence the two highest non trivial value for the
gradients.

\subsubsection{Triangular lattice} 
\begin{figure}[tbp] 
\centerline{\psfig{figure=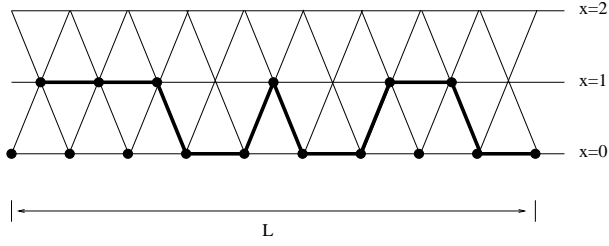,width=8cm}}
\caption{Gradient percolation on the triangular lattice with $L_g=2$ (the sites 
on the line $x=1$ are occupied with probability $1/2$). Thick disks
represent occupied sites, and the thick line joins the sites belonging
to the GP front.}
\label{trian2.fig} 
\end{figure}

\paragraph{\em Exact computations for $L_g=2$} 
We find occupied sites only on the lines $x=0$ and $x=1$. For $1\leq
i\leq L$, let $X_i$ be the random variable which has value $1$ if the
i$^{th}$ site of the line $x=1$ is occupied, and $0$ otherwise. Then,
the $X_i$'s are independent, and each takes value $1$ with probability
$1/2$ (see Fig.~\ref{trian2.fig}).
\begin{itemize}
\item
Since all the occupied sites on the line $x=1$ belong to
the front, we have $N_f(1)=\sum_{i=1}^{L} X_i$. The mean number of
points of the front on this line is thus ${\mathbf{E}}(N_f(1))=L/2$.
\item
On the line $x=0$ all the sites are occupied, but a site will belong
to the front if and only if at least one of its two neighbours on the
line $x=1$ is not occupied (see Fig.~\ref{trian2.fig}). Thus, $N_f(0)=L-\sum_{i=1}^{L} X_iX_{i+1}$, and the mean 
number of points of the front on the line $x=0$ is then
${\mathbf{E}}(N_f(0))=L-L/4=3L/4$.
\end{itemize}

Finally, we get $$n_f(0)=\frac{3}{4} \hspace{0.2cm} \mathrm{and}
\hspace{0.2cm} n_f(1)=\frac{1}{2} ,$$ and thus we can compute the mean
length $N_f$, the mean position $x_f$ and the width $\sigma_f$ of the
front, and find: $$ \frac{N_f}{L} = \frac{5}{4}=1.25 \, ,
\hspace{0.2cm} \sigma_f =
\frac{\sqrt{6}}{5} =0.4898979... \, , $$ 
$$ x_f=\frac{2}{5}=0.4 \hspace{0.2cm}
\mathrm{and} \hspace{0.2cm} p(x_f) = 1- \frac{x_f}{L_g}=0.8 \, . $$

\begin{figure}[tbp] 
\centerline{\psfig{figure=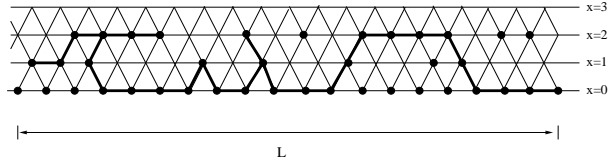,width=8cm}}
\caption{Gradient percolation on the triangular lattice with $L_g=3$ (the sites 
on the line $x=1$ are occupied with probability $2/3$, and the ones on
the line $x=2$ with probability $1/3$). Thick disks represent occupied
sites, and the thick line joins the sites belonging to the front.}
\label{trian3.fig} 
\end{figure}

\paragraph{\em Exact computations for $L_g=3$} 
For $1\leq i\leq L$, let $X_i$ be the random variable which has value
$1$ if the i$^{th}$ site of the line $x=2$ is occupied, and $0$
otherwise. Then, the $X_i$'s are independent, and each takes value $1$
with probability $1/3$. In the same way, we also define $Y_i$ as the
random variable which has value $1$ if the i$^{th}$ site of the line
$x=1$ is occupied, and $0$ otherwise. The $Y_i$'s are also
independent, and each takes value $1$ with probability $2/3$ (see Fig.~\ref{trian3.fig}).

\begin{itemize}
\item  A site on the line $x=2$ will belong to the front if it is occupied
and if it does not belong to an isolated cluster (i.e. a cluster of
occupied sites on the line $x=2$, surrounded by empty sites, see Fig.~\ref{trian3.fig}). We can write this in terms of $X_i$ and $Y_i$ in
the following way: $N_f(2) = X_1+...+X_L - \sum_{i=1}^{L} \sum_{n\geq
1} n (1-X_{i-1})X_i ...X_{i+n-1} (1-X_{i+n+1}) (1-Y_i)...(1-Y_{i+n})$.
The mean number of points of the front lying on the line $x=2$ is thus
$${\mathbf{E}}(N_f(2))=\frac{1}{3}L - L\times \frac{4}{9} \times
\frac{1}{3}
\times \sum_{n\geq 1} n \frac{1}{9^n} = \frac{5}{16}L .$$ 
 
\item 

A site on the line $x=1$ will belong to the front if it is occupied
and if: either at least one of its two neighbours on the line $x=2$ is
empty, either these two neighbours are occupied but the site is
connected to the ``sea'' on its left or on its right (see for example,
on Fig.~\ref{trian3.fig}, the third occupied site from the left on
the line $x=1$). Thus, $$N_f(1) = \sum_{i=1}^{L} Y_i (1- X_i X_{i-1})
+ \sum_{i=1}^{L} Y_i X_i X_{i-1} (S_i +S'_i - S_i S'_i), $$ where
$S_i=\sum_{n\geq 1}
(1-Y_{i-1})...(1-Y_{i-n})X_{i-2}...X_{i-n}(1-X_{i-n-1})$ is a binary
($0$ or $1$) random variable which has value $1$ when the site $i$ (on
the line $x=1$) is connected on its left to the ``sea''. $S'_i$ is the
analogous of $S_i$ for the connection on its right. We notice that
$S'_i$ and $S_i$ are independent, identically distributed, and their
mean value is ${\mathbf{E}}(S_i)={\mathbf{E}}(S'_i) =\sum_{n\geq
1}(1/3)^n (1/3)^{n-1}\times 2/3=1/4$.

The mean number of points of
the front lying on the line $x=1$ is thus
$${\mathbf{E}}(N_f(1))=\frac{2}{3}(1-\frac{1}{9})L + \frac{2}{3}\times
\frac{1}{9} (2\times \frac{1}{4} - \frac{1}{4^2})L = \frac{5}{8}L .$$ 
 
\item 
A site on the line $x=0$ will not belong to the front if: else, its
two neighbours on the line $x=1$ are occupied, or else, one of these
two neighbours is empty but the site is not connected to the ``sea''
(this is for example the case of three consecutive sites of the line
$x=0$ on the right of Fig.~\ref{trian3.fig}). Thus, we have:
$N_f(0)= L - \sum_{i=1}^{L}Y_i Y_{i+1} - \sum_{i=1}^{L} \sum_{n\geq 1}
(n+1) Y_i(1-Y_{i+1})...(1-Y_{i+n})Y_{i+n+1} X_i X_{i+1}...X_{i+n}$.
Then, the mean number of points of the front lying on the line $x=0$
is $${\mathbf{E}}(N_f(0))=L - \frac{4}{9}L - L \times
\frac{4}{9}\times
\frac{1}{3}\sum_{n\geq 1} (n+1)  \frac{1}{9^n} = \frac{223}{432} L .$$ 
\end{itemize} 

To summarize, we get $$n_f(0)=\frac{223}{432} , \hspace{0.2cm}
n_f(1)=\frac{5}{8} \hspace{0.2cm}
\mathrm{and} \hspace{0.2cm} n_f(2)=\frac{5}{16} ,$$ and then we can compute 
the mean length $N_f$, the mean position $x_f$ and the width
$\sigma_f$ of the front, and find: $$ \frac{N_f}{L} =
\frac{157}{108}=1.453704... \, \, , \sigma_f =
\frac{9\sqrt{335}}{157\sqrt{2}}=0.7419084... \, \, , $$ 
$$ x_f=\frac{135}{157} =0.8598726... \hspace{0.2cm}
\mathrm{and} \hspace{0.2cm} p(x_f) = 0.7133758...  \,\, .$$ 
 
\subsubsection{Square lattice} 
The situation for the square lattice is a little different from the
situation for the triangular lattice, because of the problem of the
choice between 4-connexity and 8-connexity. In our case, the
definition of the front will be the following: the front is the set of
occupied sites which are connected through occupied 4-neighbours to
the line $x=0$, and have an empty 8-neighbours which belongs to the
empty 8-connected component of the line $x=L_g$.

\paragraph{\em Exact computations for $L_g=2$} 
The computations for $L_g=2$ in the case of the square lattice are
very similar to the ones in the triangular case. All the occupied
sites of the line $x=1$ belong to the front, and a site on the line
$x=0$ belongs to the front if at least one of is three neighbours on
the line $x=1$ is empty.  Thus, we get $$n_f(0)=1-\frac{1}{2^3} =
\frac{7}{8} \hspace{0.2cm} \mathrm{and}
\hspace{0.2cm} n_f(1)=\frac{1}{2} ,$$ and then we can compute the mean length 
$N_f$, the mean position $x_f$ and the width $\sigma_f$ of the front:
$$ \frac{N_f}{L} = \frac{11}{8}=1.375 \, \, , \sigma_f =
\frac{2\sqrt{7}}{11}=0.4810457... $$ 
$$ x_f=\frac{4}{11}=0.36363636.... \hspace{0.2cm}
\mathrm{and} \hspace{0.2cm} p(x_f)=0.81818181....\, \,  .$$

\begin{figure}[tbp] 
\begin{center} 
\begin{tabular}{lr} 
\psfig{figure=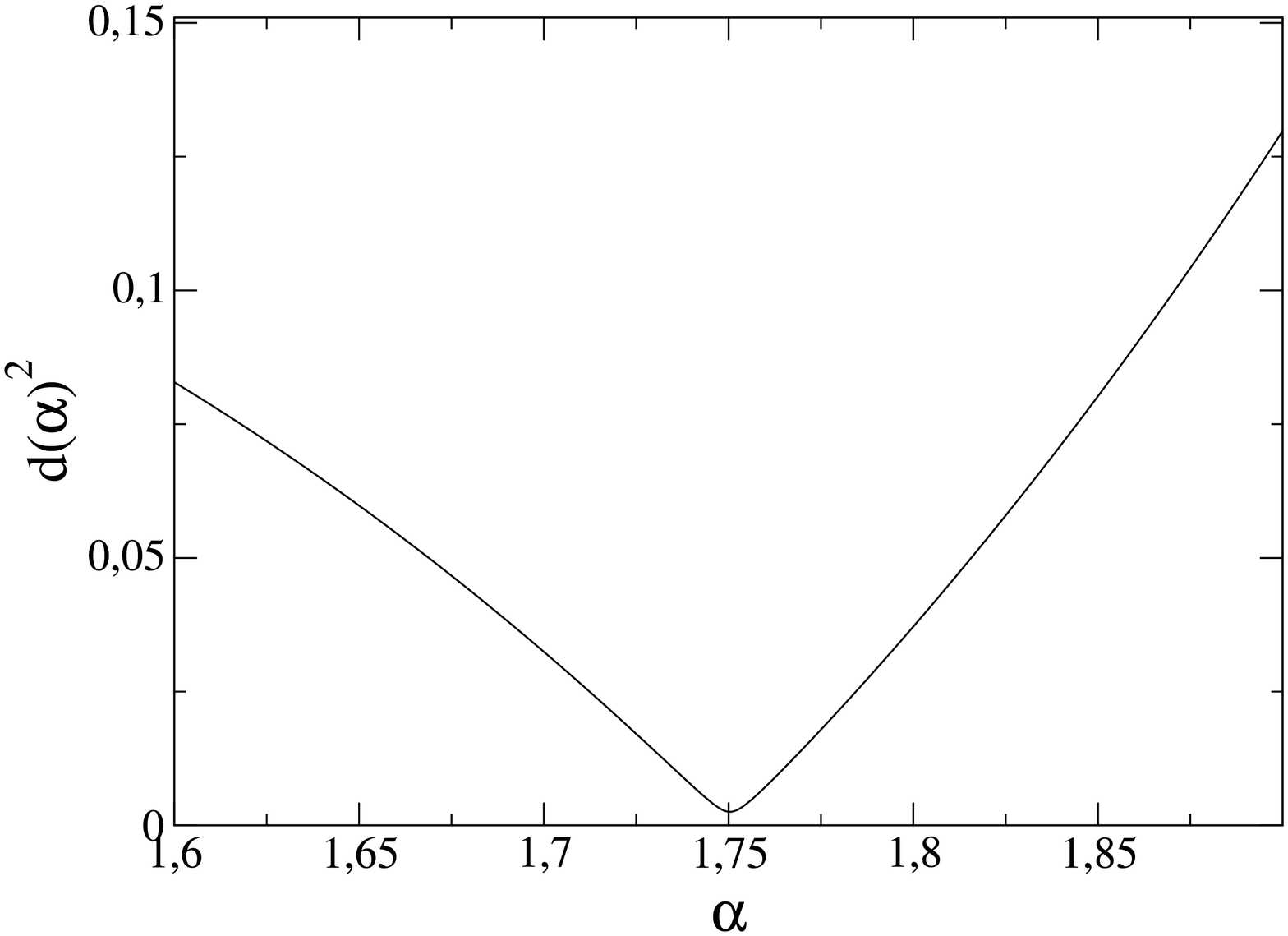,width=5.5cm}&
\psfig{figure=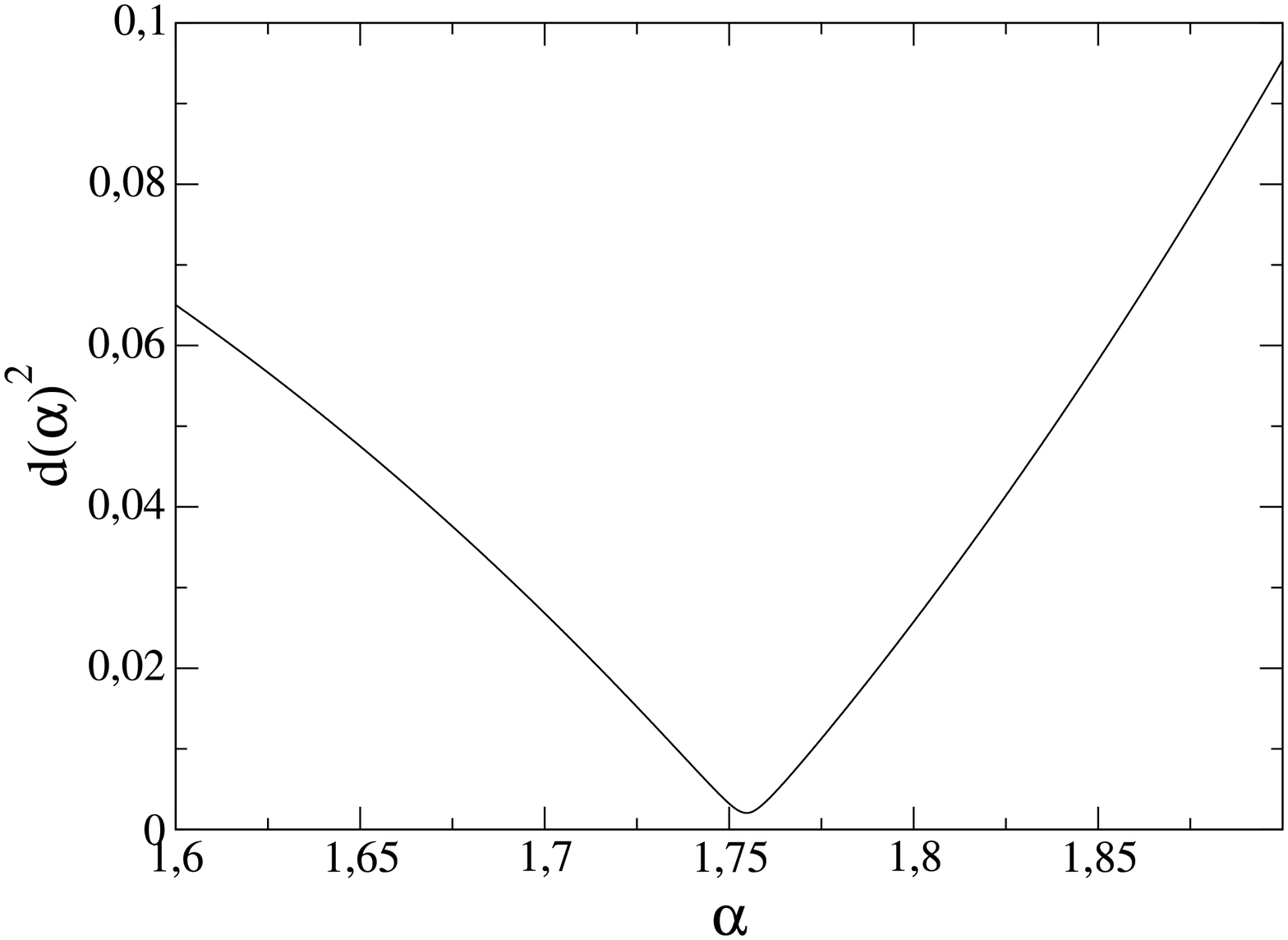,width=5.5cm}
\end{tabular} 
\end{center} 
\caption{Determination of the best exponent value: Left, square lattice: 
$\alpha=1.750$; Right: triangular lattice, $\alpha=1.754$.}
\label{fig:1.75best} 
\end{figure} 
 
\paragraph{\em Exact computations for $L_g=3$}
For $L_g=3$, the computations are similar to the ones in the case of
the triangular lattice, but a little more complicated since a site has
more neighbours, and thus more different geometric configurations have
to be considered.  But it can be done, and we finally get:
$$n_f(0)=\frac{8029}{11664} , \hspace{0.2cm} n_f(1)=\frac{47}{72}
\hspace{0.2cm} \mathrm{and} \hspace{0.2cm} n_f(2)=\frac{13}{48} ,$$ and for 
the mean length $N_f$, the mean position $x_f$ and the width
$\sigma_f$ of the front, we find: $$ \frac{N_f}{L} =
\frac{9401}{5832}=1.611968... \, \, , \sigma_f =
\frac{9\sqrt{576049}}{9401}=0.7266046....$$ 
$$ x_f=\frac{6966}{9401}=0.7409850...
\hspace{0.2cm} \mathrm{and} \hspace{0.2cm} p(x_f)= 0.7530050.... \, \, .$$

\begin{figure}[tbp] 
\begin{center} 
\begin{tabular}{cc}
\psfig{figure=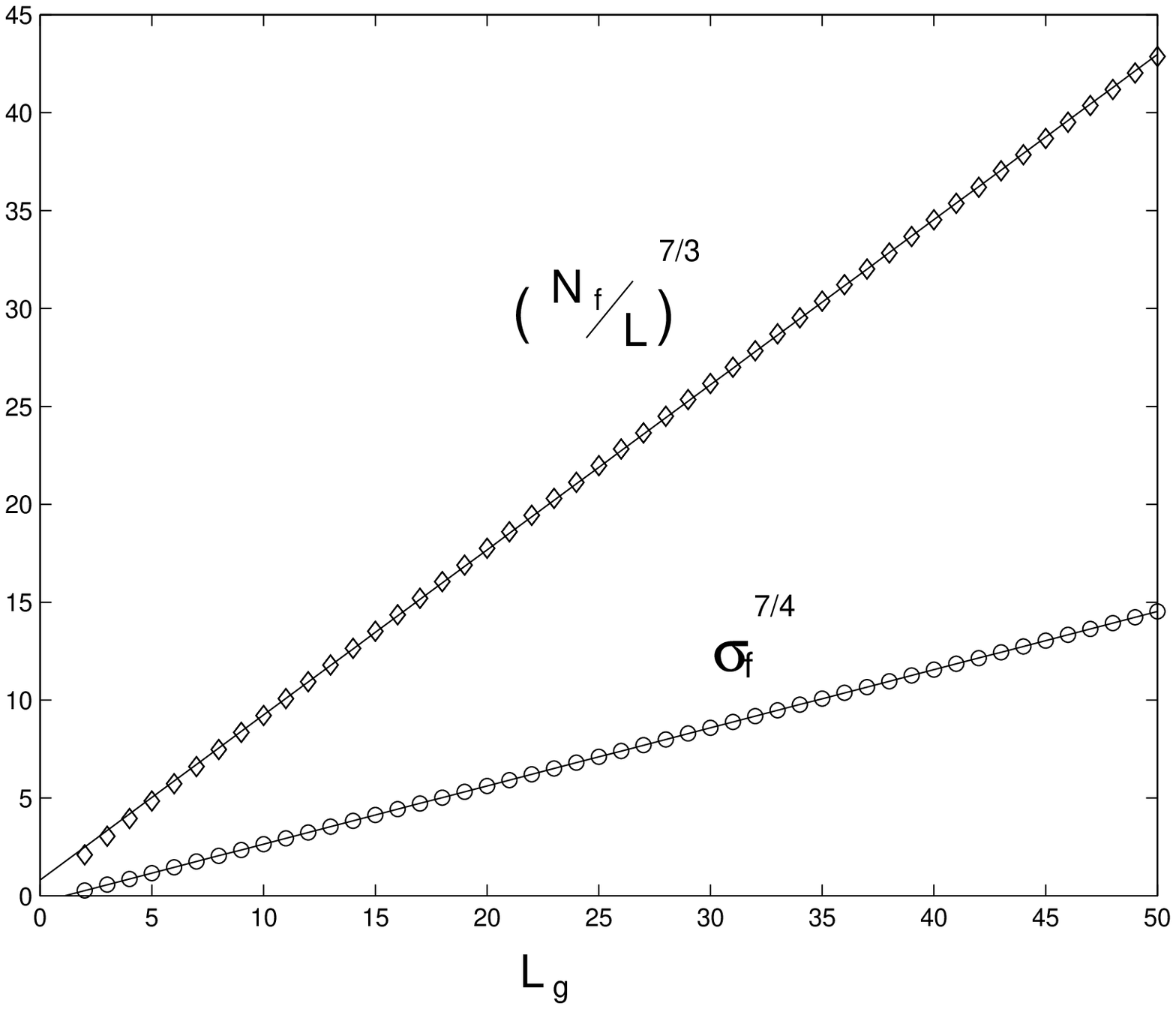,width=6cm} & 
\psfig{figure=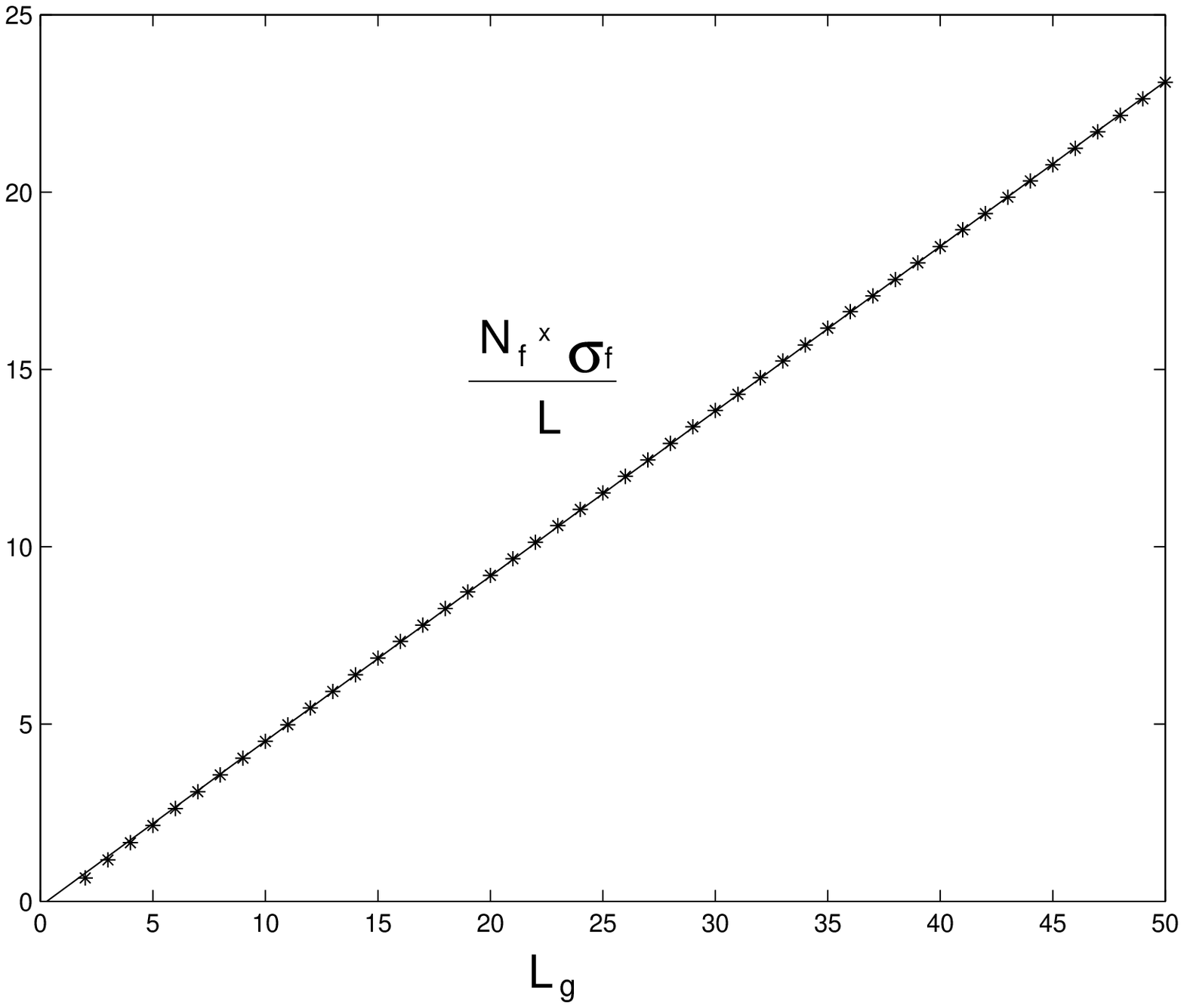,width=6cm} 
\end{tabular} 
\end{center} 
\caption{ Numerical results for the square lattice. Left: the circles resp. diamonds 
represent respectively $\sigma_f^{7/4}$ and $(N_f/L)^{7/3}$. Right:
the stars represent $N_f\times\sigma_f/L$. The lines are the ones
obtained by linear fit on the data for $L_g=4$ to $50$.}
\label{fig:larg17etNf} 
\end{figure} 
 
\subsection{Numerical results} 
The problem is then to compare the numerical GP laws to these exact
values. As will be shown, the numerical results verify the above power
laws with such precision that the question arises of the existence of
a simple mathematical power law extending from $L_g=1$ to infinity. To
try to answer this question we proceed in two steps. First we test these
laws on the numerical results obtained for $L_g$ between 4 and 50 for
the square and triangular lattices by searching the best numerical
power laws followed by the width. Considering arbitrary exponents
$\alpha$ between $1.6$ and $1.9$, we study $\sigma_f^{\alpha}$ as a
function of $L_g$ between $4$ and $50$. For each $\alpha$ value, there
is a best line $\sigma_f^{\alpha}=a_{\alpha}(L_g+b_{\alpha})$ fitting
the numerical $\sigma_f^\alpha$. The introduction of the term
$b_{\alpha}$ is justified by the fact that when a power law is
verified for large systems it includes always the possibility that a
small (as compared to the system size) term could contribute but in a
negligible manner. But here the size itself is small or very small. On
the other hand, one should remark that for $L_g=1$ the width is
strictly 0 so that some negative value of $b_{\alpha}$ should be
present. In the next step, the mean error $d(\alpha)$, defined by
$d(\alpha)^2=(1/47)\sum_{L_g=4}^{50}
(\sigma_f(L_g)^\alpha-a_{\alpha}(L_g+b_{\alpha}))^2$, is measured
numerically as a function of $\alpha$. The results are shown in
Fig.~\ref{fig:1.75best}. There is a clear minimum for $\alpha \approx
1.75$, showing that this exponent gives the best power law fit. Once
the best fit with the empirical data is made one has the best values
for the parameters $a$ and $b$: $a=0.297$ and $b=-1.094$. Note that
$b$ should be strictly equal to $-1$ in order to obtain a null width
for the trivial case $L_g=1$.

\begin{table} 
\begin{center} 
\begin{tabular}{|c|c|c|c|} 
\hline  \, {\bf square lattice} \, & \, $L_g=1$ \, & \, $L_g=2$ \, & \,
$L_g=3$ \, \\  
\hline  \, exact $\sigma_f$ \, & $0$ & $0.4810$ &  $0.7266$ \\
\hline  \,  $\sigma_f$ (4-50) \, & $-0.13$ & $0.47$ & $0.72$ \\  
\hline  \, $\sigma_f$ (4-5) \, & $-0.106$  &  $0.478$  &  $0.726$ \\  
\hline  \, $\delta\sigma_f$ (4-5) \, & $0.025$  &  $0.006$  &  $0.002$ \\  
\hline  \, exact $N_f/L$ \, & $1$ & $1.3750$ &  $1.6120$ \\  
\hline  \,  $N_f/L$ (4-50) \, & $1.24$ & $1.48$ & $1.68$ \\  
\hline  \,  $N_f/L$ (4-5) \, & $1.109$  & $1.393$  &  $1.615$ \\  
\hline  \, $\delta(N_f/L)$ (4-5) \, & $0.017$  & $0.009$  &  $0.004$ \\  
\hline 
\end{tabular} 
\medskip 
\caption{Comparison between exact and extrapolated results for the square
lattice. The data $(4-50)$
resp. $(4-5)$ correspond to extrapolated values from the respective ranges
$(4-50)$ resp. $(4-5)$ (see text). $\delta$ is the confidence interval.\label{table1} 
} 
\end{center} 
\end{table} 

\begin{table} 
\begin{center} 
\begin{tabular}{|c|c|c|c|} 
\hline  \, {\bf triangular lattice} \, & \, $L_g=1$ \, & \, $L_g=2$ \, & \,
$L_g=3$ \, \\  
\hline  \, exact $\sigma_f$ \, & $0$ & $0.4899$ &  $0.7419$ \\
\hline  \,  $\sigma_f$ (4-50) \, & $-0.15$ & $0.48$ & $0.74$ \\  
\hline  \, $\sigma_f$ (4-5) \, & $-0.151$  &  $0.480$  &  $0.740$ \\  
\hline  \, $\delta\sigma_f$ (4-5) \, & $0.019$  &  $0.006$  &  $0.002$ \\  
\hline  \, exact $N_f/L$ \, & $1$ & $1.2500$ &  $1.4537$ \\  
\hline  \,  $N_f/L$ (4-50) \, & $1.16$ & $1.36$ & $1.46$ \\  
\hline  \,  $N_f/L$ (4-5) \, & $1.014$  & $1.262$  &  $1.457$ \\  
\hline  \, $\delta(N_f/L)$ (4-5) \, & $0.015$  & $0.008$  &  $0.004$ \\  
\hline 
\end{tabular} 
\medskip 
\caption{Comparison between exact and extrapolated results for the triangular
lattice. The data $(4-50)$
resp. $(4-5)$ correspond to extrapolated values from the respective ranges
$(4-50)$ resp. $(4-5)$ (see text). $\delta$ is the confidence interval.\label{table2} 
} 
\end{center} 
\end{table} 

Another verification of the extreme GP power laws can be obtained
from the study of the front length or of the quantity $(N_f/L)^{7/3}$
as a function of $L_g$. In Fig.~\ref{fig:larg17etNf} (left plot), the diamonds
represent the values of $(N_f/L)^{7/3}$ and the best linear fit has
equation $Y=c(L_g+d)$ with $c=0.843$ and $d=0.959$. This shows indeed
that the exponents $4/7$ and $3/7$ can be used down to the steepest
gradients for which the frontier is no more fractal. 

We also study
the quantity $N_f\times\sigma_f/L$ as a function of $L_g$: the result
is shown on Fig.~\ref{fig:larg17etNf} (right plot). The best linear fit (on the
values from $L_g=4$ to $50$ has equation $Y=\tilde{c}(L_g+\tilde{d})$
with $\tilde{c}=0.465$ and $\tilde{d}=-0.278$.

One can then extrapolate the $\sigma_f$ values to the case $L_g=1$,
$2$ and $3$. The results are given in Table~\ref{table1} for both
lattices. One observes that the numerical extrapolations correspond to
the exact values with an apparent good precision. However no firm
conclusion can be drawn without discussion of the numerical
uncertainties. The values shown on Fig.~\ref{fig:larg17etNf} are
averaged over $100$ trials on a length $L=5.10^{5}$. As the fit occurs
through a power law it is difficult to give a confidence interval for
the coefficients $a$ and $b$ (obtained from a least square linear
regression on the values of $\sigma_f^{7/4}$).

In order to obtain a better control on the numerical precision of $a$
and $b$ we made extensive computations of the two cases $L_g=4$ and
$L_g=5$ with 100 trials on a length $L=5.10^{5}$. Doing so we obtain
the mean values with their standard deviation:
$\sigma_f(4)^{7/4}=0.8658\pm 0.0009$ and for
$\sigma_f(5)^{7/4}=1.1610\pm 0.0013$.
Thus if we compute the equation of the line
$\overline{a}(L_g+\overline{b})$ which interpolates the two points
$(4,\sigma_f(4)^{7/4})$ and $(5,\sigma_f(5)^{7/4})$, we obtain
$\overline{a}=0.2952\pm 0.0022$
and $\overline{b}=-1.066\pm 0.041$.
This last result shows that the value $-1$ is compatible with
$\overline{b}$ and its statistical error. Given the numerical values
for $L_g=4$ and $5$, we can also get extrapolated values for
$\sigma_f$ for $L_g$ smaller, together with their confidence
interval. The result (see Tables~\ref{table1} and~\ref{table2}) is that the predicted values are
very close to the exact ones.  For $(N_f/L)^{7/3}$, in the same way we
obtain a linear interpolation of the values for $L_g=4$ and $5$, with
coefficients $\overline{c}=0,893\pm 0.014$ and $\overline{d}=0.427\pm
0.074 $.

A comparison of the main fitting results for the two geometries
(square and triangular) is shown in Table~\ref{table3}. 
  
\begin{table} 
\begin{center} 
\begin{tabular}{|c|c|c|} 
\hline   & \, {\bf square lattice}  \, & \, {\bf
triangular lattice} \, \\ 
\, {\bf numerical results} \,  & $100\times 500000$ \, & \, $100\times 500000$ \\ 
\hline  \, $\sigma_f^{7/4}$ fit for \, & $a= 0.297$,  &
$a= 0.315 $,  \\
data $Lg=4\rightarrow 50$ & $b= -1.094 $ & $b= -1.113 $ \\ 
\hline  \, $(N_f/L)^{7/3}$ fit for \, & $c= 0.843$, &
$c= 0.636$, \\ 
data $Lg=4\rightarrow 50$ &  $d= 0.959 $ &  $d= 1.214$ \\
\hline  \, $\sigma_f\times N_f/L$ fit for \, & $\tilde{c}= 0.465$,
 & $\tilde{c}= 0.426$,  \\ 
data $Lg=4\rightarrow 50$ & $\tilde{d}= -0.278$ & $\tilde{d}= -0.198$ \\
\hline  \, exact linear fit from \, &
$a= 0.2940$, & $a= 0.3062$, \\ 
 $\sigma_f^{7/4}(2)$ and $\sigma_f^{7/4}(3)$ &  $b= -1.0548$ & $b= -1.0632$ \\
\hline 
\end{tabular} 
\end{center}

\caption{ Numerical results of the linear fit for the
square and triangular lattices. The data are obtained from the
averaged over 100 trials with a length $L=500000$.\label{table3}}
\end{table} 

\subsection{Extreme Gradients for EGP}

To check the extreme gradient regime for EGP numerical simulations have been performed for large systems ($L\ge 5000$) in a wide range of etchant volumes $V$. As discussed in~\cite{GBS}, the ratio
$L/V$ is proportional to the self-established gradient generated by the
dynamical process, at least in the small gradient regime. Here we show
how it is  possible to check the whole range of gradients (from small
to extreme) without explicitly knowing the exact value of the gradient. 

Let us assume that there exists such a parameter $g$, or equivalently
$L_g=1/g$, such that $\sigma_f \propto L_g^{4/7}$ and $N_f \propto
L_g^{3/7}$. In such a case, one should have, along the lines discussed
above: 
\begin{equation} 
\sigma_f^{7/4} = A \left(\frac{N_f}L\right)^{7/3} + B
\label{linear-extreme}
\end{equation}

\begin{figure}[tbp] 
\centerline{\psfig{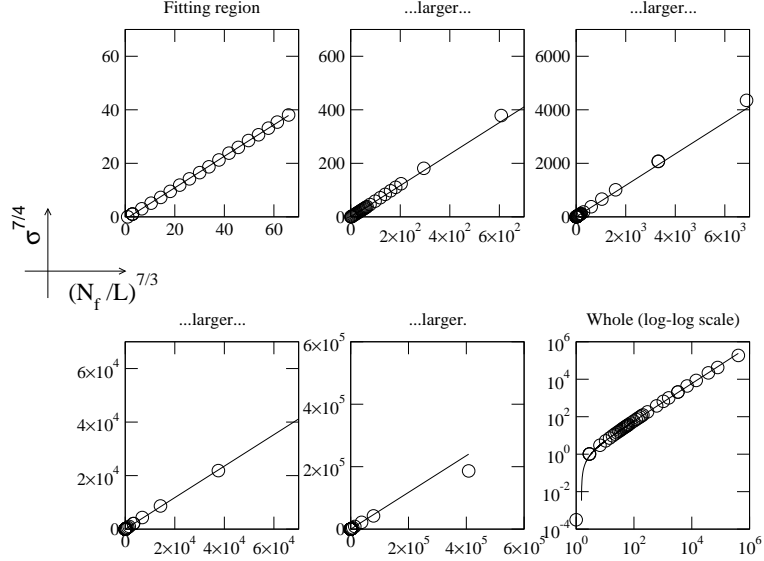}}
\caption{ E.G.P numerical scaling extrapolated from the extreme gradient regime. In the first graph
(top-left) the data are fitted against a linear relation. Then the
predicted values for larger ranges are compared with simulations. The
last graph (bottom-right) displays the whole range of data,
in a log-log plot.}
\label{extreme-EGP} 
\end{figure}

In order to check the validity of the above relation on the entire
gradient range, we proceeding in the following manner.
First, we use relation~(\ref{linear-extreme}) to determine the parameter $A$ and $B$
fitting the data on an extreme gradient (non fractal) range.
Once the parameters $A$ and $B$ have been determined, we use, we use
relation~\ref{linear-extreme} to check the agreement with the
numerical results obtained for small gradients, in the  fractal
regime.
In Fig.~\ref{extreme-EGP} the outcome of such
a procedure is shown, in the case of a triangular lattice.
In the first graph (top left) the fitting region is shown. The data
correspond to the simulation of more than $100$
realizations of  systems sized $L\ge 10000$ for $V$ ranging from
$V=L$ to $V\approx 3L$. The resulting values of $N_f/L$ correspond, in
the GP model, to $L_g$ ranging approximatively from $1$ to $70$.
The result of the fit is $A=0.58804$ and $B=-0.89278$. 
The following four graphs (from left to right in the first and then in
the second rows) display the values predicted by the linear
relation~(\ref{linear-extreme}) (straight line) compared with the
simulation results (for a system sized $L=5000$). Each graph
corresponds to a scale ten times larger than the previous graph. Finally, the
last graph (bottom right) shows the whole range of simulation results together with
the linear relation~(\ref{linear-extreme}), in a log-log scale.

\subsection{Comparison between Gradient Percolation and Etching Gradient Percolation}

\begin{figure}[tbp] 
\centerline{\psfig{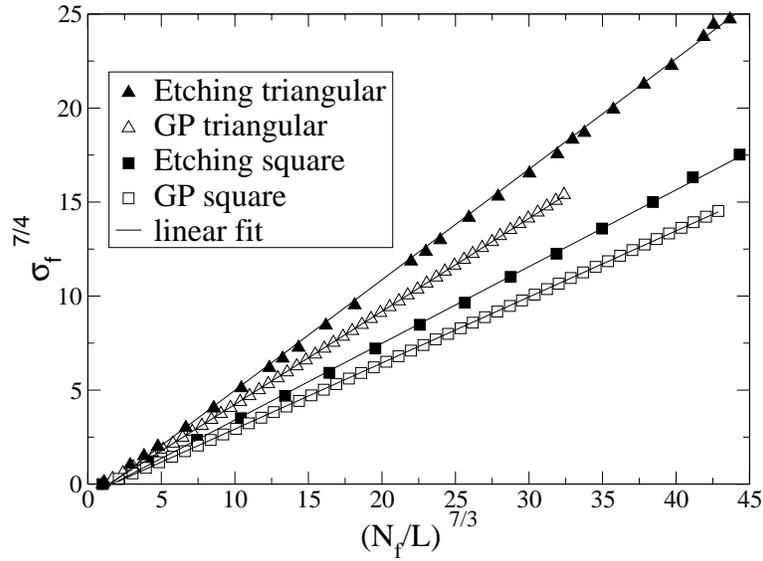}}
\caption{Extreme Gradients: GP (empty symbols) versus EGP (filled
symbols) for square and triangular geometries.}
\label{extreme-GP+EGP} 
\end{figure}

The process of creation of the interface being different, the question
arises of the comparison between the two geometries. For this we
compare the extreme gradient results of the two models. In
Fig.~\ref{extreme-GP+EGP} we show the numerical results for GP and EGP
for both lattices.  The result of the linear fit, with the
relation~(\ref{linear-extreme}) are displayed in Table~\ref{table-final}.

First note that, as expected, the slopes are different for the two
lattices. More interestingly, given a lattice geometry, the GP model
and the EGP model have slightly different coefficients $A$ and
$B$. This shows that GP and EGP are not strictly equivalent, even if
they belong to the same universality class (with respect to
exponents). The  differences between GP and EGP can be explained in
the following way. 
In EGP the probability of invasion by the solution is
applied on the previous {\em irregular} frontier while in GP the probability
changes row by row. In that sense in EGP the front penetration may be
slightly increased as compared to GP. 

\begin{figure}[tbp] 
\centerline{\psfig{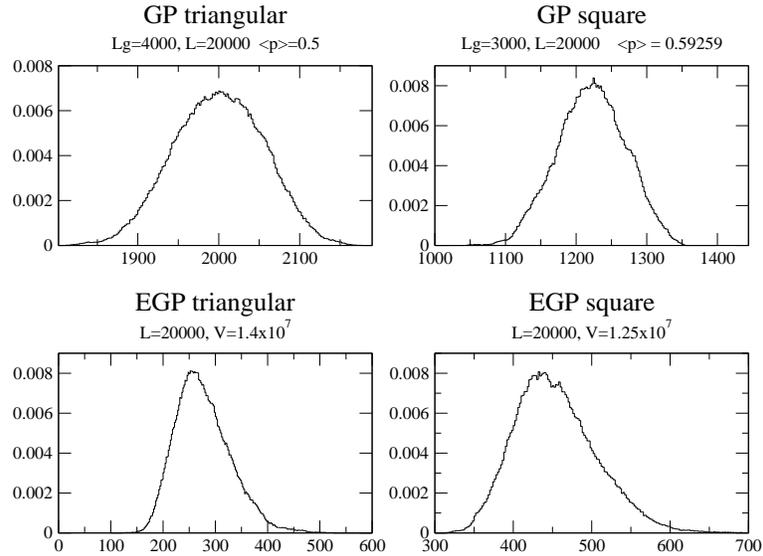}}
\caption{Depth distribution for the sites of the final front. Note the
asymmetry for the {\bf EGP} model, in contrast  to the symmetric {\bf GP} distributions.}
\label{histo} 
\end{figure}

Another quantity that displays a difference between the two models,
is the depth (i.e. x-coordinate)
distribution of the front sites, as shown in Fig.~\ref{histo}. 
Note that, at odds with the GP case, the EGP is asymmetric.
To understand such an asymmetry one has to recall the mechanism that
creates a gradient in EGP.
There, at each time $t$ there is an etching power $p(t)$ and the front
is at an average distance $x(t)$. This means that $p$ depends
implictly on $x$, through the dummy variable $t$. Consequently,
there is an effective ``self-generated'' gradient, but there is no reason why this
gradient should be constant as it is the case for standard GP. 
In particular in the last period of etching, $p$ varies only weakly
below $p_c$~\cite{GBS} while the penetration still increases. This means
that the gradient corresponding to the last period is smaller,
inducing a wider distribution for the front. 
Alternatively if GP was studied with a concave $p(x)$ (decreasing
gradient) the frontier would naturally be asymmetric.

\begin{table}  
\begin{center}  
\begin{tabular}{|c|c|c|}  
\hline  \, {\bf square lattice} \, & \, $A$ \, & \, $B$ \\
\hline  \, EGP \,& $0.40808$  & $-0.67458$ \\ 
\hline  \, GP \,& $0.35142$ & $-0.58746$ \\ 
\hline  \, {\bf triangular lattice} \, & \, $A$ \, & \, $B$ \\
\hline  \, EGP \,& $0.58804$  & $-0.89278$ \\ 
\hline  \, GP \,& $0.49459$  & $-0.70487$   \\ 
\hline  
\end{tabular}  
\medskip  
\caption{Comparison between GP, EGP. The table shows
the results of the fit performed with the relation
(\protect{\ref{linear-extreme}}) on the data shown in
Fig.~\protect{\ref{extreme-GP+EGP}}\label{table-final}  
}
\end{center}  
\end{table}

\specialsection{Discussion and Conclusion}

We have seen above that the detailed study of the extrapolation from
the best numerical values corresponding to $L_g = 4$ or $L_g = 5$ was,
for the square lattice, compatible with the exact values within
numerical uncertainties. At this point one could conclude that the
numerical results are compatible with the existence of a single
mathematical law for the width dependence, this law being valid from
$L_g=1$ to infinity. Note that the quality of the random number
generator could intervene in the purely numerical results and
uncertainties, but when we tested the case $L_g = 2$ or $L_g = 3$ we did obtain
numerically the exact results. Still, as long as a mathematical proof has not
been given, it is not possible to conclude on the exact values of the
coefficients $a$ and $b$.

The question of a unique mathematical power law can also be addressed 
from the exact results only. One can note
that, remarkably, each time one tries to fit an observable $y$ (as
$(N_f/L)^{7/3}$ or $\sigma^{7/4}$) using a function of $L_g$ like:
\[
y=a(L_g+b)
\]
the values of the term $b$ in the fit of the width are close to $-1$
(see Table~\ref{table3}). If the above law exists from $L_g=1$ to infinity, it
suggests that the real value of $b$ is exactly -1 as the width is null
in the trivial case $L_g=1$. As we have exact values, one can compute
the equation of the line $y=a(L_g+b)$ defined by the two points
$(2,\sigma_f(2)^{7/4})$ and $(3,\sigma_f(3)^{7/4})$. One obtains
$a=0.294$ and $b=-1.055$. These values are close to the values
obtained from the above numerical fit ($a=0.297$ and $b=-1.09$), and
here again $b$ is close but not equal to $-1$.
Why is there a small mismatch with the simplest law? 
The answer to this question is two-fold. 

\begin{enumerate}
\item
It is possible that the exact power law is not valid for $L_g=1$ or both for $L_g=1$ and $2$. These cases
could be "anormal" as in these cases the perimeter and the accessible perimeter are the
same (i.e. there is no  Grossman-Aharony effect~ \cite{Grossman}).
It is then possible that these two cases  cannot be explained by the same mathematical
law.

\item
The small discrepancy
could be related to the fact that  a slightly different definition of the
interface, leading to very similar results for large values of $L_g$
(where universality enters), can give slightly different
exact values for $L_g=1,2,3$.
Our frontier definition considers
only the occupied sites. It gives to these sites a privilege role
whereas one should also consider the frontier of the empty cluster. In
fact this is not new in percolation studies \cite{Rosso1,Ziff87} where
it was shown that the barycenter between the frontier of the occupied
cluster and the frontier of the empty cluster was a more natural
object. It notably permitted better computations of the percolation
threshold.  We have studied the statistical width of the local
barycenter which can also be computed exactly for $L_g=2$ or $3$. The
results show the same behavior as described above i.e. a value of $b$
close but not equal to $-1$. The question remains then open to define
the nature of the geometrical object which would really display a $b$
value exactly equal to $-1$. The answer to that question would
certainly have interesting consequences for the SP problem itself.

\end{enumerate}

In summary, it has been shown that the classical power laws of
gradient percolation can be extended to extreme gradients with the
same fractal exponents although the systems present no fractal
geometry. Several comments can be drawn on these results.

First, from the purely mathematical point of view, our results suggest
that there exists a conservation law which stipulates that the length
of the correlated frontier is strictly proportional to the gradient
length. This hypothesis remains to be proved rigorously.

Secondly, in a wider theoretical frame, the fact that the exponents
4/7 and 3/7 are valid down to the smallest $L_g$ values (or the steepest
gradients) suggests that these exponents play the same type of role
here that the exponent 1/2 intervening in the fluctuations of the sum
of independent identical random variables. In that last case the
exponent applies to {\it any} number of random variables starting from
1, 2 or 3 up to infinity.  The exponents 4/7 and 3/7 may then play a
more important role than critical exponents which only exist in the
thermodynamic limit.

Third, the fact that the same exponent has been found for the square
and the triangular lattice (and in two different models) marks a
universal behavior that, in principle, is quite unexpected here. Up to
now, the main physical argument to explain universality, i.e. the
independence of exponents from the microscopic details of the model,
was based on the long range of the correlations which appear in the
proximity of a critical point. In the extreme gradient case, on the
contrary, the (geometrical) correlations of the interface 
are so small that one cannot justify the observed universality.

Finally, our study suggests an intrinsic method to determine whether a given
rough interface belongs to gradient percolation, without knowledge of
the gradient and whatever the width of the interface.  This can be
very helpful  to understand the properties of diffused contacts
between materials.

\end{document}